\documentclass[12pt,preprint]{aastex}
\usepackage{graphicx}
\usepackage{subfigure}

\newcommand{\igr}{\objectname{IGR J16358-4726}}
\newcommand{\sgr}{\objectname{SGR 1627-41}}
\newcommand{\chan}{{\it Chandra}}
\newcommand{\integral} {{\it INTEGRAL}} 
\newcommand{\xmm}{{\it XMM-Newton}} 
\newcommand{\sax}{{\it BeppoSAX}} 
\newcommand{\asca}{{\it ASCA}} 
\newcommand{\cps}{count~s$^{-1}$}
\newcommand{\be}{\begin{eqnarray}}
\newcommand{\ee}{\end{eqnarray}}

\newcommand{\fluxunits}{ {\rm erg~cm}^{-2} {\rm s}^{-1} }

\slugcomment{draft version: \today}

\shorttitle{\igr}
\shortauthors{Patel et al.}

\begin{document}

\title{A Possible Magnetar Nature for \igr}

\author{S.~K.~Patel\altaffilmark{1,2}, 
J.~Zurita\altaffilmark{3}, 
M.~Del Santo\altaffilmark{4}, 
M.~Finger\altaffilmark{2}, 
C.~Kouveliotou\altaffilmark{1}, 
D.~Eichler\altaffilmark{5},
E. G\"{o}\u{g}\"{u}\c{s}\altaffilmark{6}
P.~Ubertini\altaffilmark{4}, 
R.~Walter \altaffilmark{3}, 
P.~Woods\altaffilmark{2}, 
C.~A.~Wilson\altaffilmark{4}, 
S.~Wachter\altaffilmark{7}, 
A.~Bazzano\altaffilmark{4}
}

\altaffiltext{1}{NASA/Marshall Space Flight Center, National Space Science and Technology Center, VP62, 320 Sparkman Drive, Huntsville, AL 35805}
\altaffiltext{2}{Universities Space Research Association, 6700 Odyssey Dr. NW, Suite 203, Huntsville, AL, 35806}
\altaffiltext{3}{\integral\ Science Data Centre, 16 Ch. d'{\'E}cogia, 1290 Versoix, Switzerland; Observatoire de Gen{\'e}ve, Chemin des Maillettes 51, 1290 Sauverny, Switzerland}
\altaffiltext{4}{INAF/Istituto di Astrofisica Spaziale e Fisica cosmica - Roma, via del Fosso del Cavaliere 100, 00133, Roma, Italy}
\altaffiltext{5}{Physics Department, Ben$-$Gurion University, Be'er$-$Sheva 84105, Israel}
\altaffiltext{6}{Sabanc{\i} University, FENS, Orhanl{\i}$-$Tuzla 34956 {\.I}stanbul/ Turkey}
\altaffiltext{7}{Spitzer Science Center, California Institute of Technology, MS 220-6, Pasadena, CA 91125}

\begin{abstract}
We present detailed spectral and timing analysis of the hard x-ray transient \igr\ using multi-satellite archival observations.  A study of the source flux time history over 6 years, suggests that lower luminosity transient outbursts can be occuring in intervals of at most 1 year. Joint spectral fits of the higher luminosity outburst using simultaneous \chan/ACIS and \integral/ISGRI data reveal a spectrum well described by an absorbed power law model with a high energy cut-off plus an Fe line. We detected the 1.6 hour pulsations initially reported using \chan/ACIS also in the \integral/ISGRI light curve and in subsequent \xmm\ observations.  Using the \integral\ data we identified a spin up of 94 s ($\dot P = 1.6\times 10^{-4}$), which strongly points to a neutron star nature for \igr. Assuming that the spin up is due to disc accretion, we estimate that the source magnetic field ranges between 10$^{13} \sim 10^{15}$ G, depending on its distance, possibly supporting a magnetar nature for \igr.
\end{abstract}


\section{Introduction}

\igr\ belongs to a new class of transient hard X-ray sources discovered with ESA's INTErnational Gamma-Ray Astronomy Laboratory \cite[\integral;][]{Win03}. Kuulkers (2005) describes the spectral and temporal properties of ten of these sources discovered during regular scans of the Galactic plane with the Imager on-board INTEGRAL \citep[IBIS;][]{Ube03} through the end of 2004. They all exhibit very high intrinsic photo-absorption (N$_{\rm H}$ $\gtrsim$ 10$^{23}$ cm$^{-2}$) resulting in hard spectra with power law spectral indices $0.5 <\Gamma< 2.1$. Four of these new transients exhibit long period coherent pulsations in their high energy (X- and $\gamma-$ray) fluxes in the range of 228 s to $\sim5900$ s; so far they have been interpreted in the literature as High Mass X-ray Binaries (HMXBs), based on their {\it spectral} similarities to other X-ray pulsars in these systems \citep{Kuu05,Lut05}. In only one of these sources, however, an orbital period of $8.9\pm0.1$~d has also been found, supporting an HMXB nature for the source \citep[IGR J16320-4751;][]{Cor05}. 

Thus far, \igr\ stands out as the sole outlier among these new \integral\ X-ray transients, having one of the hardest X-ray spectra and the longest period of all. The source was discovered with \integral\ by \cite{Rev03} when it went into outburst on 2003 March 19. Our contemporaneous \chan\ observations led to the discovery of large amplitude coherent intensity modulations in the source light curve with $P=5880$ s \citep{Kou03,Pat04}. \cite{Pat04} showed that the source spectrum is extremely hard ($\Gamma=0.5$) and highly absorbed (N$_{\rm H}$ $=1.3\times10^{23}$ cm$^{-2}$), both during the outburst and in quiescence. Radio follow up observations taken $\sim25$ days and $\sim134$ days after the outburst resulted only in flux upper limits of 0.9 mJy (4.8 GHz; Patel et al. 2004) and 7.5 mJy (0.61 GHz; Pandey et al. 2006), respectively. \cite{Wac06} report an accurate location of the source derived by off-axis PSF modeling of the initial \chan\ observation.  Their infrared observations, taken during the decaying tail of the source outburst, reveal a counterpart candidate for \igr\ different than the one discussed by \citet{Kou03} and \citet{Ami06}; their data allow both for a Low Mass X-ray Binary (LMXB) or an HMXB system scenario (based only on the brightness of the counterpart). The candidate suggested by \citep{Ami06} was observed during source quiescence; based on the lack of variability in the light curve of this putative counterpart,  \citet{Ami06} supported an HMXB nature of the source. Finally, based on the \integral\ survey observations, \igr\ is thought to be located in the Norma arm of the Milky Way, at distance ranges of $5-6$ or $12-13$ kpc, depending on which crossing of the arm with the line of sight is chosen \citep{Lut05}.  

To date, there is no conclusive evidence for the nature of the \igr\ period. \cite{Pat04} suggested that the source is either a transient accreting pulsar with an unusually long spin period, or a transient Low Mass X-ray Binary (LMXB) with a short orbit. Unraveling the nature of \igr\ is essential, as this source provides an excellent avenue towards understanding the new class of \integral\ objects. Here, we present a study of the source flux history and spectra using combined X- and $\gamma-$ray observations with \asca, \sax, \integral, \chan\ and \xmm. We describe the \integral\ and \xmm\ observations and data reduction in Section 2. In Section 3 we construct the long term flux history of \igr\ spanning $\sim 6$ years and search for evidence of possible orbital modulations; further we perform a timing analysis using \integral\ data to identify changes in the observed period. In Section 4 we perform both a wide band joint spectral analysis to better constrain the integrated source spectrum and phase resolved spectroscopy with \chan\ to search for spectral variations and features. A discussion of our results and possible models on the nature of the source is presented in Section 5.

\section{Observations and Data Reduction}
\subsection{\integral}

A significant portion of the observing time allocated to the \integral\ Core Program is dedicated to the study of hard X- and $\gamma$-ray sources in the Galaxy.  This time is mainly divided in Galactic Plane Scans (GPS) and Galactic Center Deep Exposures (GCDE). The observing strategy consists in pointings lasting 2.2 and 1.8 ks for GPS and GCDE, respectively, that are spread in various grid patterns to ensure the best use of the coded-mask instruments. As \igr\ is located near the galactic plane, it has been regularly monitored by \integral\ \citep{Lut05}. Hereafter, we will reference the \integral\ observations by {\em revolution} (Rev) number (1 total revolution is completed every $\sim 3$ days) and by individual {\em science window} (scw) number (1 scw $\sim 2$ ks). \igr\ was discovered as a transient source in Rev 52 and faded beyond detection a few days later. Since then, no other detection of this source has been reported with \integral.

We have searched the \integral\ public database to identify additional periods of \igr\ activity; our search included all Guest Observer data obtained from the start of the mission until 2003 Oct 16.3 (MJD 52928.3), as well as Core Program data obtained until 2004 Dec 2 (MJD 53341.1). Our data set comprises four epochs when the Galactic center was deeply observed resulting in a total exposure time of 4.3 Ms. The average off-axis angle for all \igr\ observations is larger than $10^{\circ}$. We have, therefore, only used data taken with the large FOV coded-mask instrument IBIS/ISGRI \citep[20~keV-1~MeV,][]{Ube03,Leb03} on board \integral\ in this investigation. Table \ref{tabDataSet} presents the details of the source visibility periods, and, for each of them, the fraction of the time when the source was effectively observed in the IBIS partially-coded field of view (PCFOV) of $29\degr \times 29 \degr$. Revolutions with net source exposure $< 1$~ks were discarded since these typically correspond to pointings during which the source crossed the PCFOV far from the center. These omitted revolutions are: 56, 61, 63, 164, 169, 179, 183, 241, 243 and 244.

The data from all pointings with the source within the ISGRI PCFOV were reduced with the Offline Scientific Analysis (OSA) software\footnote{http://isdc.unige.ch/} (v4.2 and v5.0) provided by the \integral\ Science Data Center \citep{Cou03}. This resulted in $\sim 2000$ science windows distributed between revolutions 30 and 234. For each pointing we extracted sky images and we then built mosaic images with longer exposures.  As the source is a transient, we extracted and inspected light curves from each individual pointing as well as from the composite images. During the times the source was detected we extracted light curves on shorter time scales of $\sim 50$~s. We also generated individual spectra for each pointing and then summed them to obtain one average spectrum per revolution where the source was detected. The same technique was used to generate phase-resolved spectra based on the pulse period derived with the ISGRI data. All the pointings were initially analyzed with OSA 4.2, however, the revolutions in which the source was detected were subsequently re-analyzed with OSA 5.0 (see details in Table~\ref{tabRevDetected}).  Lightcurves were corrected to the solar system barycenter using the OSA 5.0 {\sl barycent} tool.	
	
It was practically impossible to observe the source with the X-ray instrument JEM-X, which has a FOV with a diameter of $13.2^{\circ}$; moreover, it was very difficult to analyze sources lying more than $3^{\circ}$ away from the center of the JEM-X FOV.  In the single pointing where the source was located $\sim 2.5^{\circ}$ from the center (revolution 54), we do not detect it with a 3$\sigma$ upper limit of 0.005 count~s$^{-1}$. Finally, data from the third instrument on board \integral, the Gamma Ray Imaging Spectrometer (SPI), were also not analyzed because the instrument angular resolution of $2^{\circ}$ is too large for the crowded region where \igr\ is located (e.g., 4U 1630-47 is located only $19\arcmin$ away).

\subsection{\xmm}
\label{sec_xmmobs}

We observed the location of \igr\ with \xmm\ starting on 2004 February 15.6~UT and again on 2004 September 4.8~UT for $\sim 32$~ks in each pointing. The field of \igr\ was observed again by \xmm\ during a scheduled observation of \sgr\ on 2004 September 23.3~UT.  During the latter observation, the EPIC-PN was operating in Small Window Mode thus no PN data of \igr\ were collected.  \igr\ was detected $9\arcmin.8$ off-axis with the MOS detectors during this observation  
    
The observation data files were processed using the \xmm\ Standard Analysis System (SAS v6.5.0) tools {\it emchain} and {\it epchain} and HEASOFT/FTOOLS (v6.0.2).  We first identified and removed times of extremely high background rates in the PN and both MOS detectors  ($>15$~counts~s$^{-1}$ across the entire detector) and then repeatedly applied a $3~\sigma$ clipping criterion to the $0.5-10.0$ keV count rates until all the remaining events were within $\pm3~\sigma$ of the mean rate. The remaining useful exposure times are given in Table \ref{tabDataSet}.

For the first two observations, when the source was detectable and imaged on-axis, we extracted source spectra from each detector using an extraction radius of $r=18\arcsec$; we used $r=30\arcsec$ in the third observation when \igr\ was imaged off-axis. We then built instrument response files using the standard SAS tools {\em arfgen} and {\em rmfgen}.  Background spectra were collected using offset regions close to the source location for each observation with a total area 15 times the source region. Finally, we grouped spectral bins to obtain at least 15 counts per bin before background subtraction in each spectrum. 

We extracted events centered on the \chan\ source location from the first \xmm\ PN observation; the source was not detected.  Using WebPimms and assuming the \chan\ spectral model ($\Gamma=0.8$, log $N_{\rm H}=23.3$; \cite{Pat04}) we derived a $3\sigma$ upper limit on the $1-10$ keV unabsorbed flux of $5.4\times10^{-14}$~ergs~cm$^{-2}$~s$^{-1}$.  The limits derived from the MOS detectors are consistent with but not more constraining than the PN measurement. During the second \xmm\ observation we clearly detected a source at the location of \igr\ with an average total PN count rate of $(1.78\pm0.09)\times 10^{-2}$ \cps\ ($1-10$ keV). We estimated a PN background rate of $(0.31\pm0.01)\times~10^{-2}$ \cps. For the MOS1 and MOS2 detectors we found total rates of $(5.5\pm0.5)\times~10^{-3}$ \cps\ and $(6.0\pm0.5)\times~10^{-3}$\cps\ and calculated background rates of $(1.06\pm 0.03)\times~10^{-3}$ \cps\ and $(1.53\pm0.04)\times~10^{-3}$ \cps\, respectively. Source and background spectra were also extracted from the off-axis data from the final \xmm\ observation.  The MOS detectors clearly detect the source with total and background rates of $(6.9\pm0.4)\times~10^{-3}$ \cps\ and $(2.89\pm0.07)\times~10^{-3}$\cps for MOS1 and $(6.1\pm0.4)\times~10^{-3}$ \cps\ and $(3.74\pm0.08)\times~10^{-3}$\cps for MOS2.  However, the poor quality of the MOS spectra in this observation did not allow for useful spectral analysis.   

\section{Results}
\subsection{Source position}

\igr\ was brightest during revolution 52 with an average count rate of 5.8$\pm$0.1 \cps\ and a detection significance of 44.7$\sigma$ (Figure 1; $20-60$ keV). We used a mosaic of that epoch to extract the \integral\ best source position at RA $= 16^{h}35^{m}52.5^{s}$ and Dec. $=-47\degr 25\arcmin 16\arcsec$ (J2000) with an uncertainty of $20\arcsec$ that includes a $10\arcsec$ systematic error due to instrument misalignment. An additional systematic error (proportional to the source significance) of $44\arcsec$ is also present from the image reconstruction \citep{Gro03}.  This position is consistent with the location initially reported by \cite{Kou03b}. A more accurate location using the second \xmm\ observation and the initial \chan\ observations is presented in a companion paper by \cite{Wac06} together with a discussion on the results of our search for an infrared counterpart for \igr.

\subsection{Source X-ray Light Curve}

Figure 2 exhibits the ISGRI light curve of \igr\ during the 2003 March outburst of the source (Revolutions 52 through 55, spanning a total of 9 days). The gaps in the Figure correspond to Revs 53 and 56, when the source was outside the IBIS PCFOV. During the source maximum (Rev 52) the average ISGRI count rate was 5.8$\pm$0.1 \cps\, corresponding to 36 mCrab. The initial peak lasted less than 2 days; the source flux fell to 3.2 \cps\ (20 mCrab) during the following 8 days and beyond detection level $<1$ \cps\ ($\lesssim 7$ mCrab) after Rev 57.

The long term flux history of \igr\ (unabsorbed $2-10$ keV) spanning $\sim$6 years was constructed by using all available pointed observations of the source and is presented in Figure \ref{fluxhist}. The archival source flux measurements with \asca\ and \sax\ prior to the 2003 outburst were originally presented in \cite{Pat04}. The $2-10$ keV flux was estimated for each X-ray instrument by assuming a spectrum similar to the one measured with \chan\ in 2003.  The \integral\ flux was extrapolated using the best fit spectral model to the ISGRI data during the outburst (Rev 52).  Using the extrapolated \integral\ discovery flux and the two \chan\ flux measurements, we find that the outburst lightcurve decays as an exponential with an e-folding time of $9.2\pm0.6$~days. The RXTE observation of \igr\ is complicated by significant X-ray emission from the Galactic ridge. Using spectral fits obtained after modeling and subtracting the Galactic ridge component \citep{Rev03}, we estimate an unabsorbed $2-10$~keV flux of $2.9\times10^{-10} \fluxunits$ on MJD 52724.4.  This flux is nearly a factor of 2 larger than the flux we derive from our CXO measurement on MJD 52722.3.  Such a large change in the source flux is diffcult to reconcile with the constant average rate observed during the 7.1~hour CXO observation.  This data point has been omitted from Figure \ref{fluxhist} for clarity. The corresponding $2-10$ keV luminosities in Figure \ref{fluxhist} were then determined by assuming an average source distance of 7 kpc. 

The detections and non-detections of \igr\ shown in Figure 3 motivated a search for orbital periods based on the assumption that this system exhibits a behavior similar to the largest class of known transient accreting pulsars, those with Be (or Oe) star companions. These systems often have series of outbursts separated by their orbital period, with the outbursts occurring near periastron passage in an eccentric orbit \citep{Bil97}.  For each one of a finely spaced grid of trial orbital periods in the range of $10-2000$ days, an attempt was made to divide the orbit into an active phase region (containing all the detections), and a quiescent phase region (with all the non-detections). Each time such a phase segregation was possible, it was considered as a candidate period. Figure \ref{orbit_search} shows the duration of the longest possible quiescent orbital phase region for each candidate period. Note that no candidate periods longer than $\sim400$ days were found. Thus, although the search did not identify any unique orbital period, it did demonstrate that a, yet undiscovered, periodic outbursting could be occurring in intervals of at most 1 year. 

\subsection{Timing analysis}

We searched for all detections of \igr\ in the public, two-year long IBIS/ISGRI database (see also Section 2) in two energy ranges: $20-60$ keV and $60-150$ keV. We then built mosaic images for each revolution and extracted light curves at each energy band where the net exposures (at the source location) were higher than 5~ks, for better statistics.  At no time was the source detected above 60~keV.  Whenever the source was not detected in one revolution (typical $3 \sigma$ flux upper limits of $\sim 0.3$ \cps), we grouped adjacent revolutions to increase sensitivity. The final net exposure times for the four resulting detection revolutions and their dates are given in Table \ref{tabRevDetected}.

Initially, we generated Lomb-Scargle periodograms \citep{Sca82} from the ISGRI light curves with the higher count rate detections (Revs 52, 54, and 55; Figure \ref{ISGRIlc}) and clearly detected significant modulations at $\sim1.65$~h (as well as its first harmonic). Further, we divided the light curve into 2 sections (Rev 52 and Rev 54$+$55) to search for changes in modulation frequency. Although we detected significant pulsations in the initial portion of the light curve (until $\sim 7\times10^4$~s after detection) with a period of $5970\pm15$~s, in the latter portion of the light curve we found moderate evidence for pulsations consistent with the previously measured frequency. In both orbit segments we detected significant power in the $\sim 3000$~s harmonic. However, due to the poor statistics during the second ISGRI interval, we were unable to conclusively determine pulse period changes exclusively within these data using the Lomb-Scargle method. Finally, we proceeded into folding the first $\sim 10^5$~s of the ISGRI light curve on the best determined \integral\ period (Figure \ref{prof}) and compared it with the \chan\ X-ray light curve acquired on 2003 March 24 folded on the same ephemeris.  There appears to be evidence of a phase shift b
etween these two profiles suggesting evolution of the pulse profile or period over a few days; however, we caution that since the profiles are in different energy bands this shift could be due to spectral evolution.   

In the following, we describe a more detailed method for the determination of the pulse ephemeris. We again determine the pulse periods separately from the data in Rev 52 and Revs 55$+$54 combined using a method that takes into consideration the properties of the data quality and the observation windows. An important aspect of this analysis was the need to correctly account for the data gaps and the large variations in rate errors, which occurred in both of these intervals. These gaps and variations were due to changes in pointing directions during the observations, causing missing data when the source was outside the coded field of view, and increasing the errors as the source moved outward within the PCFOV. For this purpose only exposure-corrected count rates in the $20-45$ keV energy band with 50 s time resolution were analyzed.

We first determined the pulse periods using a technique based on $\chi^2$ fits of the rates. To account for the gradual decrease in flux during the observations, the rates in each interval were fitted to separate linear trends, and the rates were then divided by the fit. The trend amounted to 1.0\% in Rev 52, and 31\% over the course of Revs 54$+$55. For a grid of trial pulse periods ranging from 5000~s to 7000~s, the normalized rates were then fitted to a model consisting of a repeated pulse profile, parameterized by a Fourier expansion limited to constant, fundamental and first harmonic terms. Most of the power of the pulse was found to be in the fundamental and first harmonic, so we did not include higher harmonics. The improvement in this fit ($\Delta\chi^2$) over a fit of the normalized rates to a constant is shown versus the period in Figure \ref{delta_chisq} for both time intervals. This $\Delta\chi^2$ is equivalent to the $Z^2_2$ statistic \citep{Buc83} in the case of uniformly spaced data with uniform errors and periods much smaller than the duration of the observation. The best fitted periods were $P = 5964.7\pm 12.1$~s (Rev 52) and $P = 5870.9\pm 3.4$~s (Rev 54$+$55), where the errors include only counting statistics. 

To check these results and to quantify any noise in excess of counting statistics, each analysis interval was divided into sub-intervals and the pulse profile for each sub-interval was obtained by fitting the rates with a second order Fourier expansion pulse profile model, with the period fixed to that determined above for that interval. These were then cross-correlated with the mean profile for the full interval to obtain pulse phase measurements. The deviation of these phases from those expected from the estimated period is shown in Figure \ref{pulse_phases}. Using linear fits to the pulse phases, we obtained pulse period estimates of $P=5964.9\pm12.2$~s for Rev 52, and $P=5871.3\pm3.3$~s for Revs 54$+$55 combined, consistent with the previous results. The $\chi^2$ of these fits is 4.72 and 3.47, respectively (both with three degrees of freedom). These results are summarized in Table \ref{period_tab} with the errors increased by the square root of the reduced chi-squares to account for any noise in excess of counting statistics.

In Figure \ref{pulse_periods} we show the evolution of the \igr\ pulse periods. The two \chan\ measurements agree with the second ISGRI measurement (Rev 54$+$55), but the initial ISGRI period (Rev 52) is significantly higher than the others.  The 94~s difference between the first and second ISGRI periods is significant at the 6$\sigma$ level. If this difference is due to a gradual change in the spin period of a neutron star, the mean spin period derivative between the measurements is $(-1.65\pm 0.28)\times 10^{-4}$. The last \chan\ measurement indicates that this putative spin-up trend may have ended. Unfortunately, the Rev 54$+$55 data only weakly constrain period changes; fitting the pulse phases in Revs 54$+$55 with a quadratic, results in the estimate of $\dot P = (1.9\pm 1.1)\times 10^{-4}$. Over the 18.8 pulse cycles which occur during Rev 52, the 94~s difference in period amounts to only 1760~s or 0.30 pulse cycles of phase difference. It could be possible that the pulse profile, which our analyses assume is constant, has changed sufficiently to cause such an apparent phase shift. Figure \ref{profiles_52} shows the $20-45$ keV ISGRI rates epoch-folded with a 5965 s period, for the five sub-intervals of Rev 52 used in Figure \ref{pulse_phases}. Also shown on the profile is the position of the phase reference that would have resulted if the profiles were folded with a period of 5871.3~s. Examining these alternate phase alignments of pulse profiles, we conclude that this period difference is most likely not due to profile changes.

We have also determined a pulse period using the \xmm\ EPIC-PN observation from 2004 September; in all other data sets the source was too faint to detect pulsations. We correct the photon times of arrival in the events list to the solar system barycenter and extract all events between $0.5-10.0$ keV from a slightly smaller extraction region ($12\arcsec$) as was used in the spectral analysis (see also Section 4). We performed a $Z^2_2$ test on periods ranging from $4000-7000$ sec and clearly detected the pulsation in the EPIC-PN light curve with a period of $P=5858\pm74$~s.  This period is consistent with those measured using \chan\ and \integral\ in 2003. We present the folded PN light curve in Figure \ref{xmmprof}.

\section{Spectral analysis}

\subsection{\integral/ISGRI + \chan/ACIS-S Spectral Analysis}
The phase averaged spectral fit results from the \chan\ data alone have already been presented in \cite{Pat04}.  We have, therefore, initially fitted a power law model to the \integral/ISGRI $20-200$ keV spectrum, which resulted in an unacceptable fit ($\chi^2/\nu=79/9$). A thermal blackbody (BB) model resulted in a better fit ($\chi^2/\nu=13.7/9$) with $kT_{BB}=6.2\pm0.1$~keV; however, the resulting BB source radius was only 0.15~km (assuming a source distance of 7~kpc). Subsequently, we jointly fitted the phase averaged spectra measured with ISGRI and the initial \chan\ observation, to better characterize the \igr\ spectrum in outburst.  Since these data were collected at slightly different times, we allowed the model normalizations between data sets to be free and linked all other spectral parameters. The data were well fit by a Comptonization model \citep{Tit94} with the best fit parameters given in Table \ref{tabjointspec}. For comparions to power-law models typically used in the lower x-ray bands, we have also modeled the spectrum using an absorbed power-law model with a high energy exponential cut-off (Table \ref{tabjointspec}).

On 2003 March 24, we were fortunate to have simultaneous \integral\ and \chan\ observations providing us with the only broadband high energy spectral coverage for \igr\ to date. The \integral\ data available during the time of the \chan\ observations consist of 4 science windows from which we were able to extract $\sim 4600$~s of useful exposure time. The light curves from these observations are shown in Figure \ref{jointlc}; the overall shape of the ISGRI light curve is consistent with the shape of the \chan\ light curve during the same time interval suggesting no evolution in the hard x-ray ephemeris.  We performed joint phase-averaged spectroscopy using these simultaneous observations covering approximately one entire pulse (Figure \ref{jointlc}). The data are well represented by a Comptonization model with a Gaussian line feature at 6.4 keV. The best fit parameters are given in Table \ref{tabjointspec} and the resulting spectrum is shown in Figure \ref{jointspec}. 

\subsection{\xmm/EPIC-PN Spectral Analysis}

We employed standard $\chi^2$ methods in obtaining the best fit to the \xmm\ spectral data and analyzed the spectra using the XSPEC (v11.3.2l) spectral-fitting package \citep{Arn96}.  When fitting the MOS and PN spectra simultaneously we linked the model parameters. However, since the relative flux calibration of these instruments is somewhat uncertain, we allowed the individual normalizations to be free to obtain their best fit value.  We fitted the MOS and PN ($0.5-10.0$ keV) spectra jointly to a power-law and again to a thermal BB model.  We accounted for absorption due to intervening gas and dust from the Galaxy using the Tuebingen-Boulder ISM absorption model (XSPEC model {\sl tbabs}) with the relative abundances and cross-sections from \citet{Wil00} and with He cross section from \citet{Yan98}.  An acceptable fit ($\chi^2/\nu= 30.3/38$) resulted when the data were modeled with the power law with a best fit spectral index $\Gamma=1.1^{+0.4}_{-0.3}$ and column density $N_{\rm H}=18^{+4}_{-3} \times 10^{22}$~cm$^{-2}$.  The unabsorbed ($1-10$ keV) flux measured from this model was $7.1 \times 10^{-13}$~ergs s$^{-1}$~cm$^{-2}$.  A fit using a BB model resulted in a nearly identical quality fit ($\chi^2/\nu= 30.4/38$) with $kT_{BB}=2.3^{+0.6}_{-0.2}$~keV and $N_{\rm H}=(14 \pm 3) \times 10^{22}$~cm$^{-2}$. The unabsorbed ($1-10$ keV) flux measured for this model was $4.9 \times 10^{-13}$~ergs s$^{-1}$~cm$^{-2}$.

To test the goodness of the $\chi^2$ fitting, we also fitted the unbinned data using the C-statistic \cite{Cas79} and again using the grouped data and the $\chi^2$ fitting with the weighting method described by \citet{Chu96} as implemented in XSPEC, and found consistent results.

\subsubsection{Phase Resolved Spectroscopy}

We have extracted 10 individual phase resolved spectra from the \chan\ ACIS observation of 2003 March.  We simultaneously fit the spectra using an absorbed power-law plus Gaussian line model.  The model consists of 6 parameters for each spectrum (the equivalent hydrogen column ($N_{\rm H}$), spectral power-law index ($\Gamma$), power-law normalization ($PLNORM$), Gaussian line centroid energy ($E_{line}$), width ($\sigma_{line}$), and flux ($F_{line}$)) for a total of 60 parameters.  For all fits we allow $PLNORM$ for each spectral model to aquire its best fit value (i.e. $PLNORM$ is {\it free} and {\it unlinked}).  All the remaining parameters were forced to have the same value (i.e. were {\it linked}) for each spectrum and each parameter was free to acquire its best fit value. This fit results in a statically acceptable fit with $\chi^2/\nu=562.6/542$ and best fit model parameters of $N_{\rm H}= 3.5\pm0.1 \times 10^{23}$~cm$^{-2}$, $\Gamma=0.59\pm0.09$, $E_{line}=6.38\pm0.01$~keV, $\sigma_{line}=54\pm18$~eV, and line flux of $F_{line}=2.6\pm0.3 \times 10^{-12}$~ergs cm$^{-2}$~s$^{-1}$.

To search for variations with phase of the linked parameters, we first unlinked the $N_{\rm H}$  while continuing to hold the others linked.  This resulted in significant improvements in the fit with $\Delta \chi^2=29.1$ with an F-test probability of $F_{prob}= 0.0011$.  We next unlinked  $\Gamma$ and found an additional improvement of $\Delta \chi^2=21.9$.  To test whether variations in $\Gamma$ or $N_{\rm H}$ (or both) were responsible for the majority of the fit improvement we repeat the same test this time first unlinking $\Gamma$.  When unlinking only $\Gamma$ we find a $\Delta \chi^2=41.8$ with $F_{prob}=7.5\times10^{-6}$.  Unlinking $N_{\rm H}$ at this point negligibly improves the fit with an additional $\Delta \chi^2=9.5$.  There is clear evidence for change of spectral index with phase with the spectral index ranging from $\sim 0.1-0.9$. Figure \ref{phaseres} shows the pulse profile and the evolution of the power-law index; the spectrum is harder during the peak of the pulse and softens as the rate declines.  We conclude that the evidence for varying column density with phase during this observation is marginal at best.

 We note that in \cite{Pat04}, we reported a significant decrease in the measured $N_{\rm H}$ of (by $\sim 40\%$) between this \chan\ pointing and the one 27 days later.  In subsequent spectral fits the power-law index remains unlinked.

The emission line at 6.4~keV is most likely due to Fe fluorescent emission. We tested for variation in the 6.4 keV line parameters using a similar strategy. We performed a series of fits of the spectra after unlinking each line parameter while holding the remaining parameters linked and also performed a fit will all line parameters unlinked. In all cases the improvement in the spectral fits were not significant ($\Delta \chi^2 < 8$ for 9 additional parameters).  We find no evidence for evolution of the line with pulse phase.

Finally we searched the individual spectra for additional line features.  The most significant line-like feature in a single spectrum occurred during phase 0.2-0.3 and appeared as a marginally significant emission line feature at 7.0 keV ($\Delta \chi^2=14.5$ for 2 additional free parameters and width fixed at $\sigma_{line2}=50$~eV).  However, when we considered the number of spectral phase bins searched, we concluded that the line was not statistically significant with a probability of chance occurance of $> 50\%$.  

\section{Discussion}

There are two feasible origins for the pulsations observed in the flux of \igr: binary orbital modulation or neutron star spin modulation.  Given the substantial difference of the period measured in Rev 52 of the INTEGRAL data from all later measurements, and the expected stability of an orbital period, we must either reject the Rev 52 measurement as unrepresentative of the intrinsic period, or reject the orbital modulation hypothesis. It is also possible that pulse shape evolution during Rev 52 could have resulted in this discrepant period, however this evolution would need to mimic a gradual shifting in phase by 30\% of a cycle over the course of the observation. It is not clear how this could be explained in the context of an orbital modulation. Finally, attributing the 1.65 hour modulation to an orbital period would presumably imply that the object is an LMXB making the nature of the source difficult to reconcile with its unusually hard spectrum.

Assuming the pulsations are spin induced, there are two possible explanations for the change in spin-period between the first observation and the following observations. Either the neutron star was spun-up by accretion induced torques, or the neutron star had a glitch in spin period, analogous to those seen in rotation powered pulsars. 

We first consider a scenario where the observed modulation is the neutron star spin and the abrupt change in the frequency is attributed to the sudden tranfer of angular momentum between the superfluid interior and the pulsar crust \citep[i.e., a "glitch";][]{And75, Rud76, Alp77}.  We measure a change of frequency of $\Delta \nu = (2.7\pm0.4) \mu{\rm Hz}$ and thus a fractional change in the frequency of $\Delta \nu/\nu = 1.6 \time 10^{-2}$.  This is $\sim 10^4$ times larger than the values derived from large glitches reported from radio pulsars \citep[e.g.,][]{She96, Kra03} and anomalous x-ray pulsars which have typical $\Delta \nu/\nu \sim 10^{-6}$ \citep{Hey99,Kas00,Kas03,Mor05}. To date, there has been only one substantiated report of a glitch observed from an accreting pulsar \citep[KS 1947+300, Be system;][]{Gal04}.  The fractional change in frequency measured in the KS 1947+300 glitch was $3.7\times10^{-5}$, also significantly larger than those seen in radio or anomalous X-ray pulsars, but still much smaller than observed in \igr.  We conclude, therefore, that unless the source underwent a unique and new type of glitch, never observed before, we can exclude this phenomenon as the cause of the neutron star spin variation.

In the following scenario, we will now assume that the neutron star is fed by an accretion disk during the outburst observed by INTEGRAL. Then from the measured spin-up rate and flux, and an assumed distance, $d$,  we can determine the pulsars magnetic moment. If instead the pulsar is wind fed, the spin-up at a given mass accretion rate is always less than the spin-up rate in the disk fed case, so that the magnetic moment calculated for the disk-fed case will be a lower limit for the wind fed case. The expected spin-up rate for disc accretion is \citep[see e.g.][]{Pringle72}
\begin{equation}
   \dot\nu = (2\pi I)^{-1} \eta (GM r_A)^{1/2} \dot M \label{eq_spinup}
\end{equation}
were $I$ is the neutron star's moment of inertia, $M$ its mass, $\dot M$ the accretion rate onto the neutron star, and $r_A$ the Alf\'{v}en radius for spherical accretion, 

\begin{equation}
   r_A = (2GM)^{-1/7} \mu^{4/7} {\dot M}^{-2/7} 
   \label{eq_rA}
\end{equation}

where $\mu$ is the neutron star's magnetic moment. The model dependent factor $\eta$ is expected to be near one in the ``slow rotator" regime, where the rotation rate of the neutron star is much smaller than the rotation rate of the inner disk. For example, in the model of \citet[][eqns. 10 and 11]{Ghosh79} this factor is given by $\eta = n(0) \zeta^{1/2} (= 1.39\times0.52^{1/2}=1.00) $ for a slow rotator. 

From the 94 s period change between MJD 52717.84 and 52724.39 seen in the INTEGRAL data, we infer an average spin-up rate of $\dot\nu = 4.7 \times 10^{-12}~{\rm Hz~s}^{-1}$ ($\dot P = 1.6\times 10^{-4}$). For the same time interval we estimate an average unabsorbed 2-100 keV flux of $F=7\times 10^{-10}~\fluxunits$, using flux estimates from the joint INTEGRAL-Chandra fit of $1.2\times 10^{-9}~\fluxunits$ during Rev 52, and $5.0\times 10^{-10}~\fluxunits$ at the time of the initial Chandra observation, combined with the trend of the IBIS/ISGRI 20-45 keV rates. With these values we find from equations \ref{eq_spinup} and \ref{eq_rA} the magnetic moment

\begin{equation}
   \mu = 1.8\times 10^{32} \eta^{-7/2} m_{1.4}^{3/2} R_6^{-3} I_{45}^{7/2} d_{7}^{-6}~~{\rm G~cm}^3
\end{equation}

where $m_{1.4} = M/1.4M_\odot$, $R_6 = R/10^6$cm, $I_{45} = I/10^{45}$\,gm\,cm$^2$, and $d_{7}= d/7~$kpc. Figure \ref{accretionparameters} shows the magnetic moment (and field) as a function of distance to the source assuming canonical neutron star mass and radius.

As the mass accretion rate drops, the disk inner radius (or the magnetospheric radius) $r_0 = \zeta r_A$ will expand to the co-rotation radius $r_{co} = (GM)^{1/3} (2\pi/P)^{-2/3}$, at which point the accretion will halt because material at $r_0$ which is forced to co-rotate with the magnetic field is flung-off by the centrifugal force  \citep{Illarionov75}. The transition to this ``propeller regime" should occur when the accretion flux of \igr\ decreases to

\begin{equation}
F_{min} = 1.6\times 10^{-12} \eta^{-7} \zeta^{7/2} m_{1.4}^{7/3} R_6^{-7} I_{45}^{7} d_{7}^{-14}
          ~~\fluxunits \label{eq_fmin}
\end{equation}          

\igr\ was detected in the XMM observations with a 1-10 keV flux of  $7 \times 10^{-13}~\fluxunits$. The spectrum was hard, and pulsations clearly present, both indicating that the neutron star was still accreting. This gives us a lower limit on the distance to the source of

\begin{equation}
d > 6.7 (\epsilon/0.25)^{1/14} \eta^{-1/2} \zeta^{-1/4} m_{1.4}^{1/6} R_6^{-1/2} I_{45}^{1/2}~~{\rm kpc}
\end{equation}

where $\epsilon$ is the fraction of the total flux in the 1-10 keV band.

The neutron star in \igr\ was most likely born with a much shorter period than currently observed, and has been spun down to its current state. \citet{Davies81} discussed this spin-evolution in the context of wind accreting systems, and showed that obtaining long period systems such as \igr\ requires a combination of high magnetic moment and low mass capture rates from the stellar wind. The impact of accretion disk formation on this evolution is unclear. If we suppose the source is currently near an equilibrium period, with intervals of spin-up balanced by intervals of spin-down, than an interesting question is how long are the spin-down intervals? If the source falls back into the sub-sonic propeller regime between outbursts the spin-down rate will be \citep{Davies81}

\begin{equation}
   \dot P_{prop} = 7\times 10^{-4} \mu_{30}^2 m_{1.4}^{-1} I_{45}^{-1}~{\rm s~yr}^{-1} 
   \label{eq_prop_sd}
\end{equation}

where $\mu_{30} = \mu/10^{30}~{\rm G~cm}^3$. So the time needed to recover the 94 s period decrease seen in the during the outburst is

\begin{equation}
   \tau_{recur} = 94s/\dot P_{prop} = 3.5  \eta^{7} m_{1.4}^{-2} R_6^{6} I_{45}^{-6} d_{7}^{12}~{\rm yr}       
\end{equation}

so that if the distance to the source is greater than 8.4 kpc, we are not likely to see another outburst from it in the next 30 years. The accretion disk solutions of \citet{Rappaport04} show spin-down at low mass accretion rates which scale as equation \ref{eq_prop_sd},  but with a factor of six lower rate. 
If in fact the recurrance time is this long, we consider the fact that the transient was detected in outburst for the first time during the initial year of INTEGRAL Galactic plane scans as extremely fortuitous. Therefore, if the source is detected in a similar state again within the next 3.5 years, it most likely is a transiently accreting magnetar at $\sim~7$~kpc with a field strength of $3.4\times 10^{14}$~G.  
     
\acknowledgments
SKP acknowledges support from NASA grants NNG04GH33G and DD3-4023X. MDS has been supported by the Italian Space Agency grant I/R/046/04. DE acknowledges the Israel Science Foundation, the US - Israel Binational Science Foundation, and the Arnow Chair of Theoretical Astrophysics.  We thank E. Behar for discussions on the nature of the spectral lines.


\clearpage
\begin{deluxetable}{lcccc}
\tabletypesize{\footnotesize}
\tablecaption{Log of IGR~J16358$-$4726 Observations \label{tabDataSet}}
\tablehead{
\colhead{Observatory} & \colhead{Start} & \colhead{End}   & \colhead{Revolutions} & \colhead{Exposure\tablenotemark{a}} \\
\colhead{}            & \colhead{(MJD)} & \colhead{(MJD)} & \colhead{}            & \colhead{(days)} 
}
\startdata
INTEGRAL    & 52650	 & 52743      & 30$-$60   & 22.953 \\
INTEGRAL    & 52859	 & 52919      & 100$-$119 & 32.944 \\
INTEGRAL    & 53060   	 & 53084      & 167$-$174 & 10.164 \\
INTEGRAL    & 53236	 & 53263      & 226$-$234 & 13.411 \\
Chandra     & 52722.1691 & 52722.4939 & $-$	  & 0.297	\\
Chandra     & 52750.1484 & 52750.7093 & $-$	  & 0.545	\\
XMM-Newton  & 53051.6064 & 53051.9865 & $-$	  & 0.193/0.231/0.205	\\
XMM-Newton  & 53252.7663 & 53253.1255 & $-$	  & 0.263/0.288/0.299	\\
XMM-Newton  & 53270.6780 & 53271.2782 & $-$	  &  --  /0.411/0.414	\\
\enddata
\tablecomments{All public and Core Programme ISGRI data when \igr\ was within the ISGRI PCFOV between 2003 Jan 11 (Rev 30) and 2004 Sep 15 (Rev 234) are included.}
\tablenotetext{a}{The net useful exposure times. INTEGRAL times reflect ISGRI pointed observations where IGR~J16358$-$4726 fell within the PCFOV. Times from XMM-Newton are for the EPIC PN/MOS1/MOS2 detectors.}
\end{deluxetable}

\begin{deluxetable}{lccccc}
\tabletypesize{\footnotesize}
\tablecaption{Log of {\it INTEGRAL}/ISGRI Observations with Detections of IGR~J16358$-$4726 \label{tabRevDetected}}
\tablehead{
\colhead{Revolution} & \colhead{Start} & \colhead{End}   & \colhead{Exposure} & \colhead{$\theta$\tablenotemark{a}} & \colhead{$\sigma_{det}$\tablenotemark{b}} \\
\colhead{}           & \colhead{(MJD)} & \colhead{(MJD)} & \colhead{ks}       & \colhead{deg}                                  &\colhead {}
}
\startdata
052 & 52717.3972 & 52718.7003 & 77.1  & 11.4 (6.1-18.2)	& 44.7 \\
054 & 52722.2812 & 52724.6796 & 89.6  & 10.2 (2.5-18)	& 28.6 \\
055 & 52725.0607 & 52726.4123 & 90.4  & 11.3 (5.7-14.3) & 16.3 \\
057 & 52732.5051 & 52733.6327 & 70.5  & 11.3 (5.7-11.3) & 6.1  \\
\enddata
\tablenotetext{a}{Average off-axis angle. Numbers in parenthesis indicate the range of off-axis angle. Rev 53 and 56 were not considered since the source fell outside the PCFOV.} 
\tablenotetext{b}{The detection significance is derived from the mosaiced images for each revolution (20-60 keV).} 
\end{deluxetable}

\begin{deluxetable}{llll}
\tabletypesize{\footnotesize}
\tablecaption{Pulse Periods Determined From ISGRI Data \label{period_tab}}
\tablehead{
\colhead{Revolutions} & \colhead{Epoch\tablenotemark{a}} & \colhead{Period} & \colhead{Error} \\
\colhead{}            & \colhead{(MJD)}                        & \colhead{(s)}    & \colhead{(s)}
}
\startdata
52        & 52717.84 & 5965   & 15  \\
$54-55$  & 52724.39 & 5871.3 & 3.5 \\
\enddata
\tablenotetext{a}{The weighted mean of the observation times using the inverse square rate errors as weights.}
\end{deluxetable}

\begin{deluxetable}{lccl}
\tabletypesize{\footnotesize}
\tablecaption{\integral\ \& \chan\ Joint Spectral Fit Results \label{tabjointspec}} 
\tablehead{
\colhead{Parameter} & \colhead{Value\tablenotemark{a}} & \colhead{Value\tablenotemark{b}} & \colhead{Unit}
}
\startdata
\multicolumn{3}{l}{MODEL 1: Absorbed Comptonization Plus Gaussian} \\
\hline
$N_{\rm H}$        & $27.9\pm0.9$  	    & $36\pm2$		      & $\times 10^{22}$ cm$^{-2}$ \\ 
$E_{line}$         & $6.39\pm0.01$ 	    & $6.38\pm0.02$	      & keV \\
$\sigma$           & $0.07\pm0.02$ 	    & $0.06_{-0.06}^{+0.03}$  & keV \\
$kT_{injection}$   & $2.1\pm0.1$   	    & $0.4_{-0.4}^{+3.3} $    & keV \\
$kT_{plasma}$      & $7.9\pm0.6$   	    & $8.2\pm0.5 $	      & keV \\
$\tau$             & $3.8\pm0.7$   	    & $9.5\pm1.1 $	      &     \\ 
$F_{line}$         & $2.7\pm0.03$  	    & $3.3\pm0.7 $	      & $\times 10^{-12}$ ergs cm$^{-2}$ s$^{-1}$ \\
$F_{2-10 keV}^{c}$   & $1.37_{-0.08}^{+0.05}$ & $1.8_{-0.4}^{+1.7}$     & $\times 10^{-10}$ ergs cm$^{-2}$ s$^{-1}$ \\
$R_{ACIS/ISGRI}^{d}$ & 0.45          	    & 1		              &     \\
$\chi^2/\nu$       & $352.0/324$     	    & $140.5/120$	      &     \\
\hline
\multicolumn{3}{l}{MODEL 2: Absorbed Power Law w/ High Energy Cut-Off Plus Gaussian}\\
\hline
$N_{\rm H}$        & $33\pm1$               & $34\pm2$                & $\times 10^{22}$ cm$^{-2}$ \\ 
$E_{line}$         & $6.39\pm0.01$          & $6.39\pm0.02$           & keV \\
$\sigma$           & $0.07\pm0.02$          & $0.06_{-0.6}^{+0.3}$    & keV \\
$\Gamma$           & $0.53\pm0.08$          & $0.39_{-0.08}^{+0.1}$   &     \\
$E_{cut}$          & $19_{-2}^{+1}$         & $23\pm3$                & keV \\
$E_{fold}$         & $12.2\pm0.6$           & $16\pm2.0$              & keV \\
$F_{line}$         & $2.9\pm0.03$           & $3.2\pm0.7$             & $\times 10^{-12}$ ergs cm$^{-2}$ s$^{-1}$ \\
$F_{2-10 keV}^{c}$   & $1.71_{-0.09}^{+0.02}$ & $1.62_{-0.38}^{+0.04}$  & $\times 10^{-10}$ ergs cm$^{-2}$ s$^{-1}$ \\
$R_{ACIS/ISGRI}^{d}$ & 1.12                   & 1                       &     \\
$\chi^2/\nu$       & 361/324                & 140.5/120               &     \\
\enddata
\tablenotetext{a}{ Fit results derived using all useful ISGRI and ACIS data. Errors reflect the $1 \sigma$ uncertainty.}
\tablenotetext{b}{ Fit results derived using $\sim 4.6$~ks of ISGRI and ACIS data collected concurrently. Errors reflect the $1 \sigma$ uncertainty.}
\tablenotetext{c}{ Unabsorbed flux derived from the ACIS spectra.}
\tablenotetext{d}{ $R$ is the ratio of the normalizations of the primary model component derived from the ACIS and ISGRI spectra. The model normalizations for both data sets are forced to be linked for fits to the data collected concurrently.}
\end{deluxetable}


\clearpage
\begin{figure}[ht]
\scalebox{0.8}{\plotone{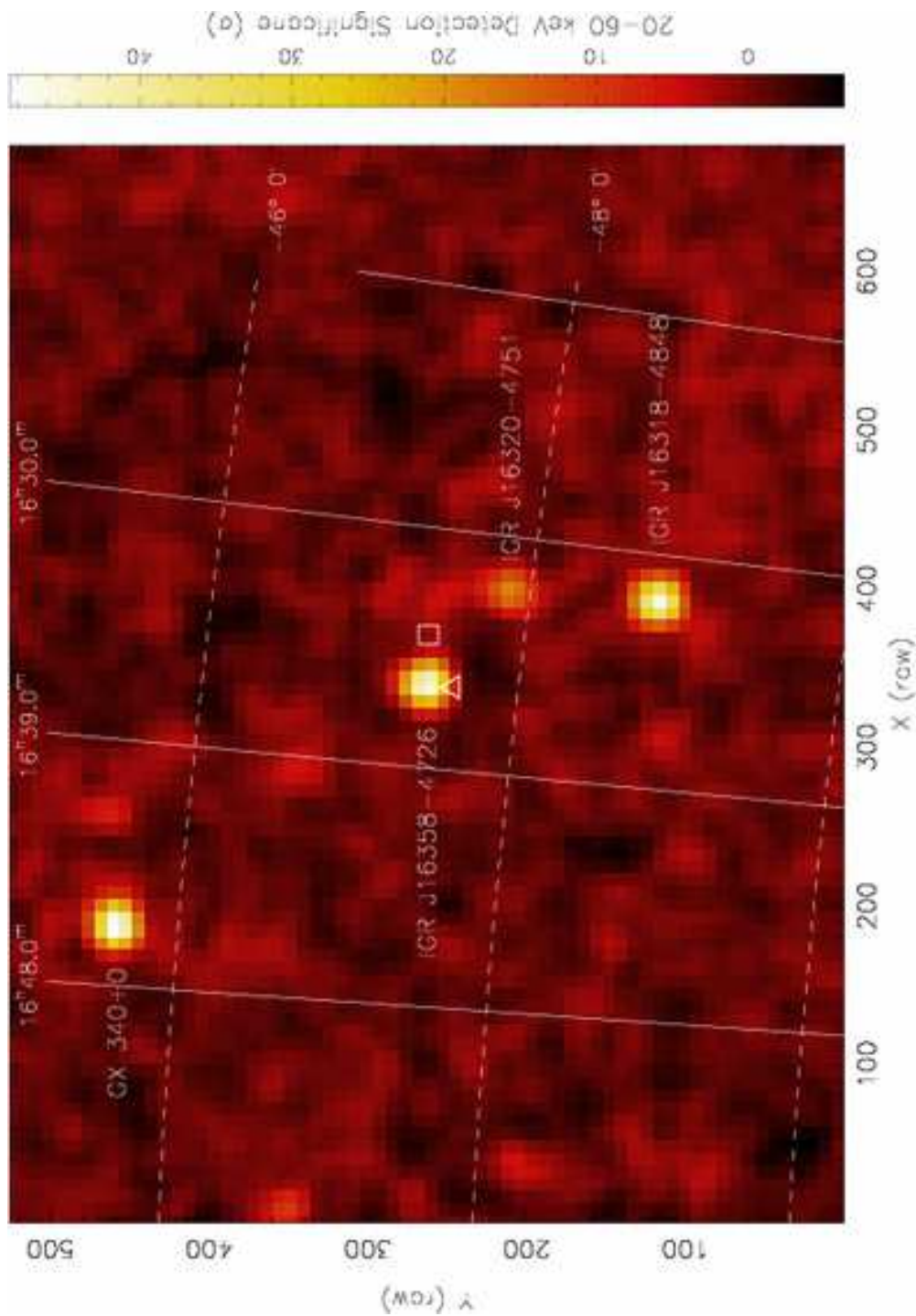}}
\caption{{\it INTEGRAL} IBIS$/$ISGRI (20$-$60 keV) image of the field containing IGR~J16358$-$4726 taken on 2003 March 19 (Rev. 52). The source detection was significant at the 44.7 $\sigma$ level in 77.1~ks of useful exposure time.  Sources detected in this portion of the sky are labeled in the image and the locations of SGR~1627$-$41 (triangle) and 4U~1630$-$47 (square) are also identified.  The latter 2 sources are not detected above 20~keV during this revolution.  For reference, the image pixel size is $4.9\arcmin$ on a side.}
\label{ISGRIimage}
\end{figure} 

\clearpage
\begin{figure}[ht]
\begin{center}
\scalebox{0.5}{\plotone{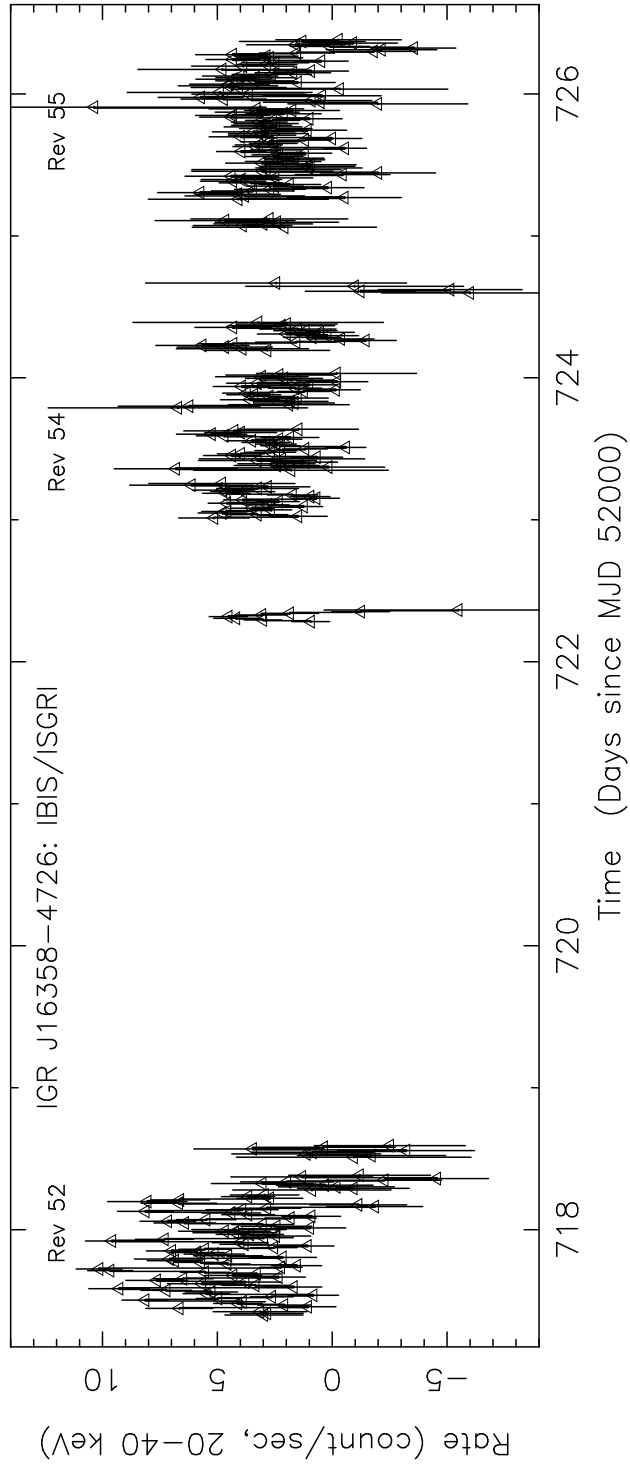}}
\end{center}
\caption{ISGRI lightcurve of IGR~J16358$-$4726 (20$-$45~keV) integrated in time bins of 1000~s, starting on 2003 March 19 and spanning $\sim 9$~days. Revs 53 and 56 were not used in this plot because the source fell outside the IBIS PCFOV (see also text and Table 2). The dashed lines indicate the time interval of the intital 26 ks Chandra observation.}
\label{ISGRIlc} 
\end{figure} 

\begin{figure}[ht]
\scalebox{1.2}{\plotone{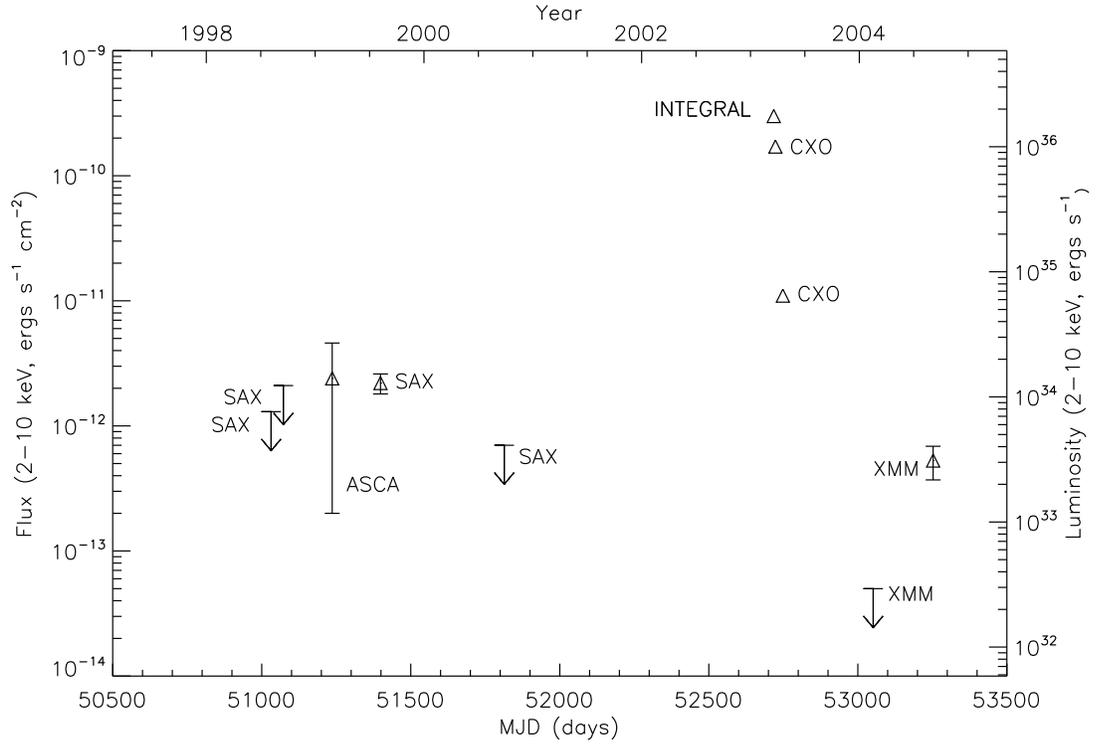}}
\caption{The unabsorbed X-ray flux history of IGR~J16358$-$4726; upper-limits at the $90\%$ significance level are indicated for 3 \sax\ measurements and 1 \xmm\ measurement.  The fluxes (left side of Y-axis) derived from the \integral\ data are extrapolated into the $2-10$ keV energy range using the best joint \chan\ - \integral\ spectral fit for Rev 52. The luminosities (right side of Y-axis) were calculated assuming an average distance of 7~kpc for the source.}
\label{fluxhist} 
\end{figure} 

\begin{figure}[!tb]
\begin{center}
\plotone{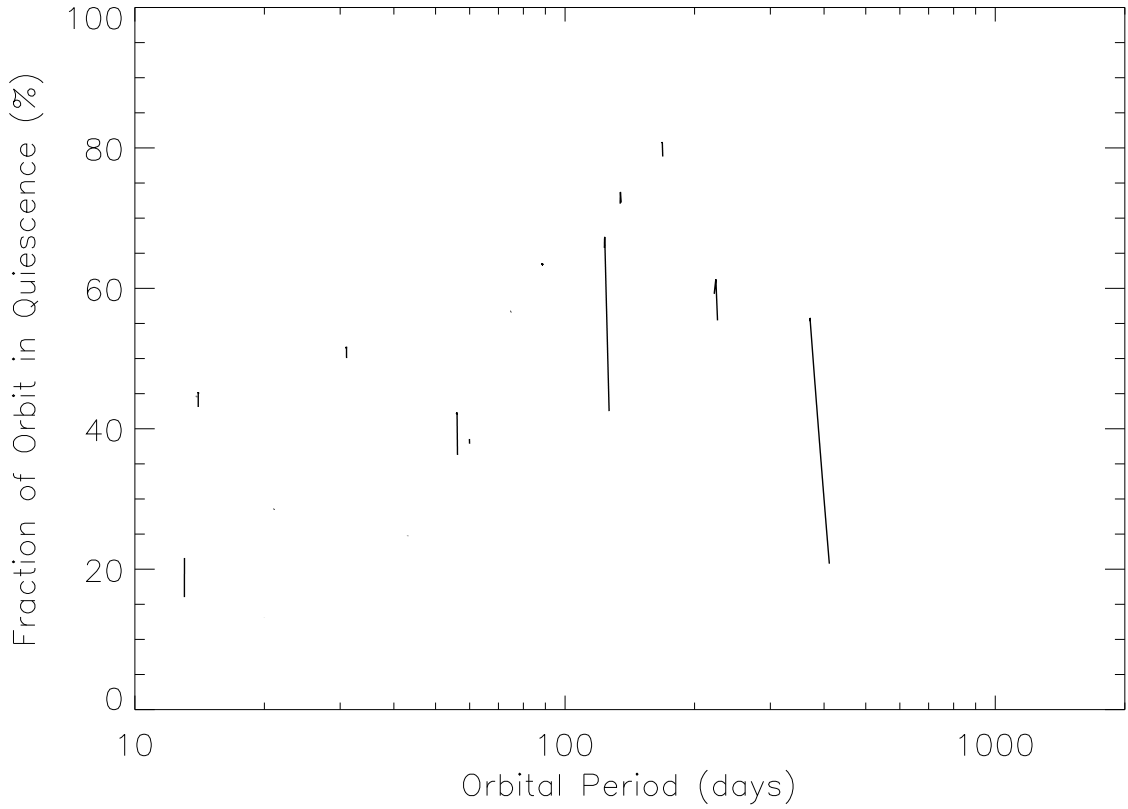}
\end{center}
\caption{Maximum duration of the quiescent orbital interval for possible orbital periods of the IGR~J16358$-$4726 system. Only the  periods are shown, where the orbit could be divided into an active phase region (containing all the detections), and a quiescent phase region (containing all the non-detections).}
\label{orbit_search}
\end{figure}

\begin{figure}[ht]   
\begin{center}
\scalebox{0.8}{\plotone{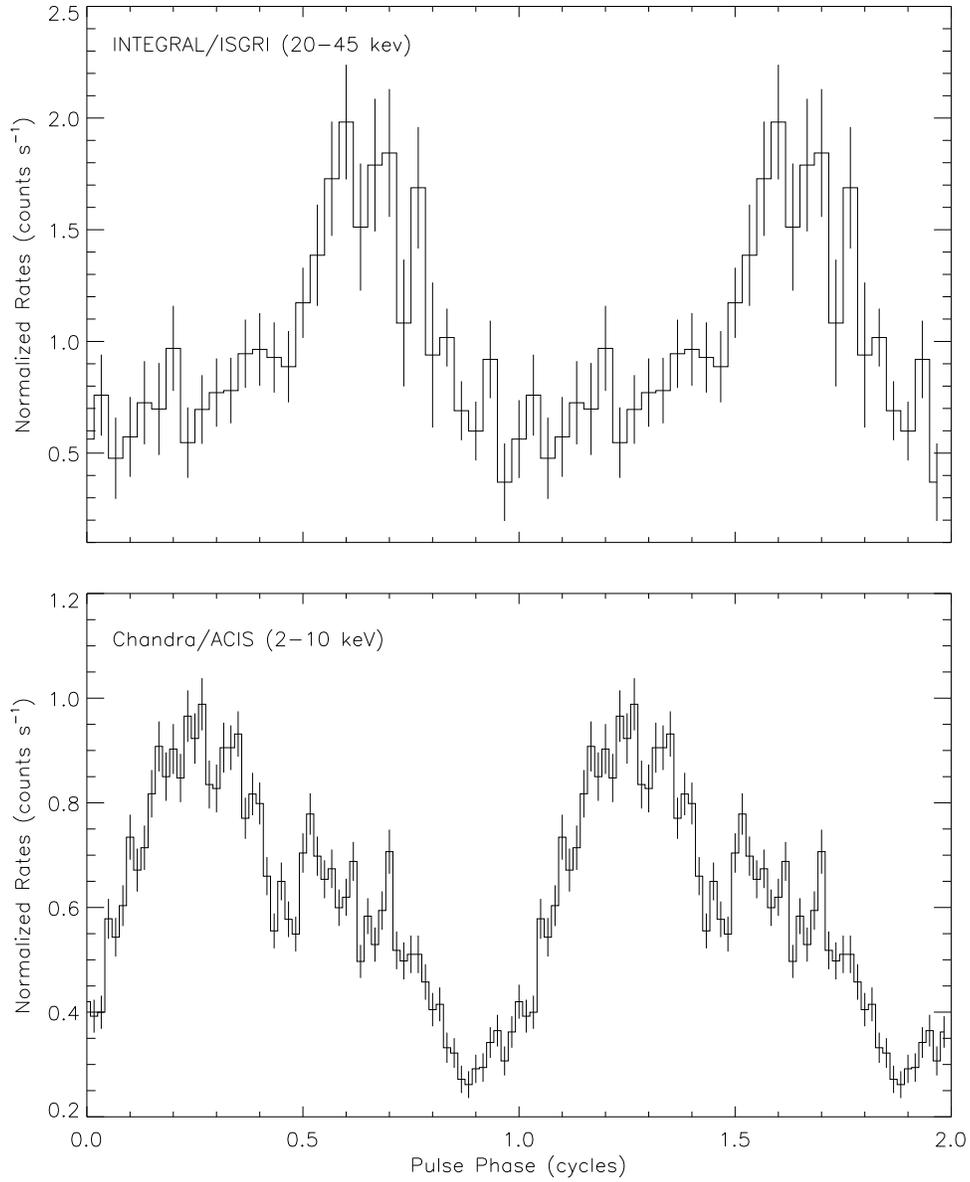}}
\end{center}
\caption{{\it Top:} \integral\ ISGRI pulse profile (20$-$45~keV) folded with a period $P=5965\pm15$~s. {\it Bottom:}  \chan\ ACIS (2$-$10~keV) pulse profile folded with the same period. Both profiles are referenced to the same date (MJD=52700).  The \chan\ data were acquired $\sim 5$~days after the ISGRI detection.} 
\label{prof} 
\end{figure} 

\begin{figure}[ht]
\begin{center}
\plottwo{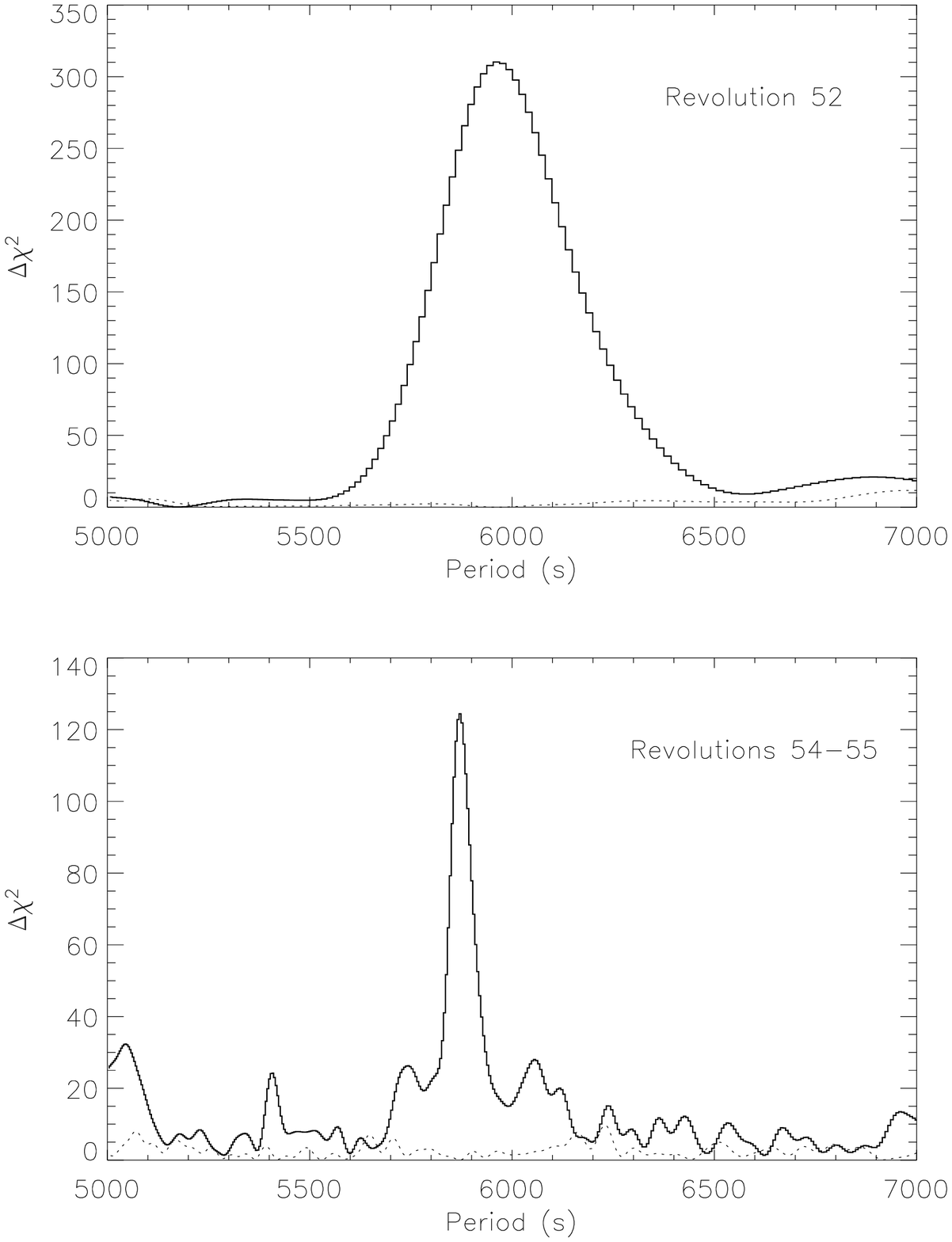}{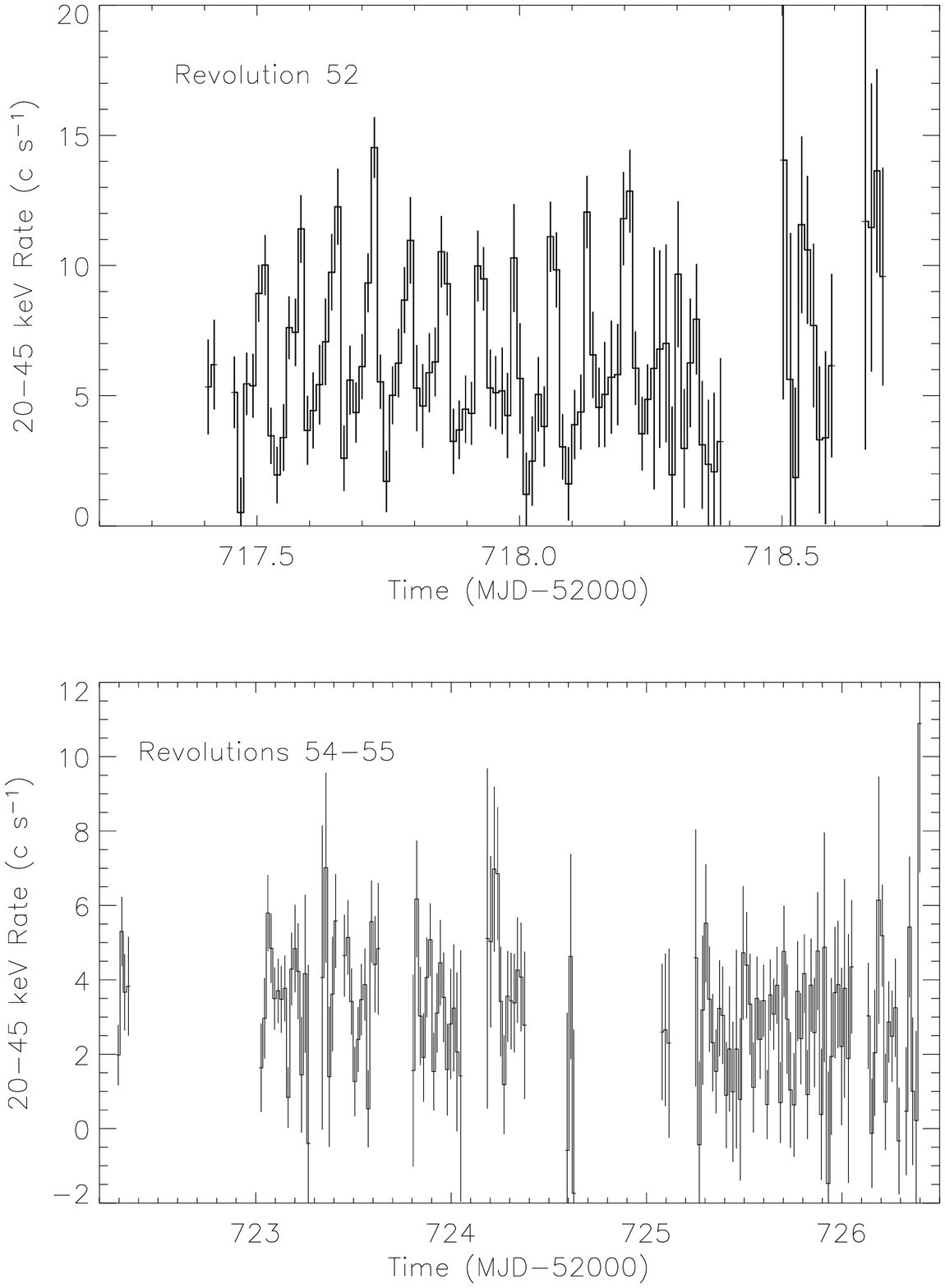}
\end{center}
\caption{The decrease in $\chi^2$ between a fit of the normalized rates to a constant and a second order Fourier expansion pulse profile is shown versus period (solid curve) for the ISGRI data in Rev 52 (top panel) and Revs 54$+$55 combined (bottom panel). Also shown (dotted curves) is the same statistic after the best fitted profile (at the peak $\Delta\chi^2$ period) was subtracted from the normalized rates. This shows that most of the power away from the main peak is caused by the pulsations at the peak $\Delta\chi^2$ period.}
\label{delta_chisq}
\end{figure}

\begin{figure}[ht]
\begin{center}
\scalebox{0.8}{\plotone{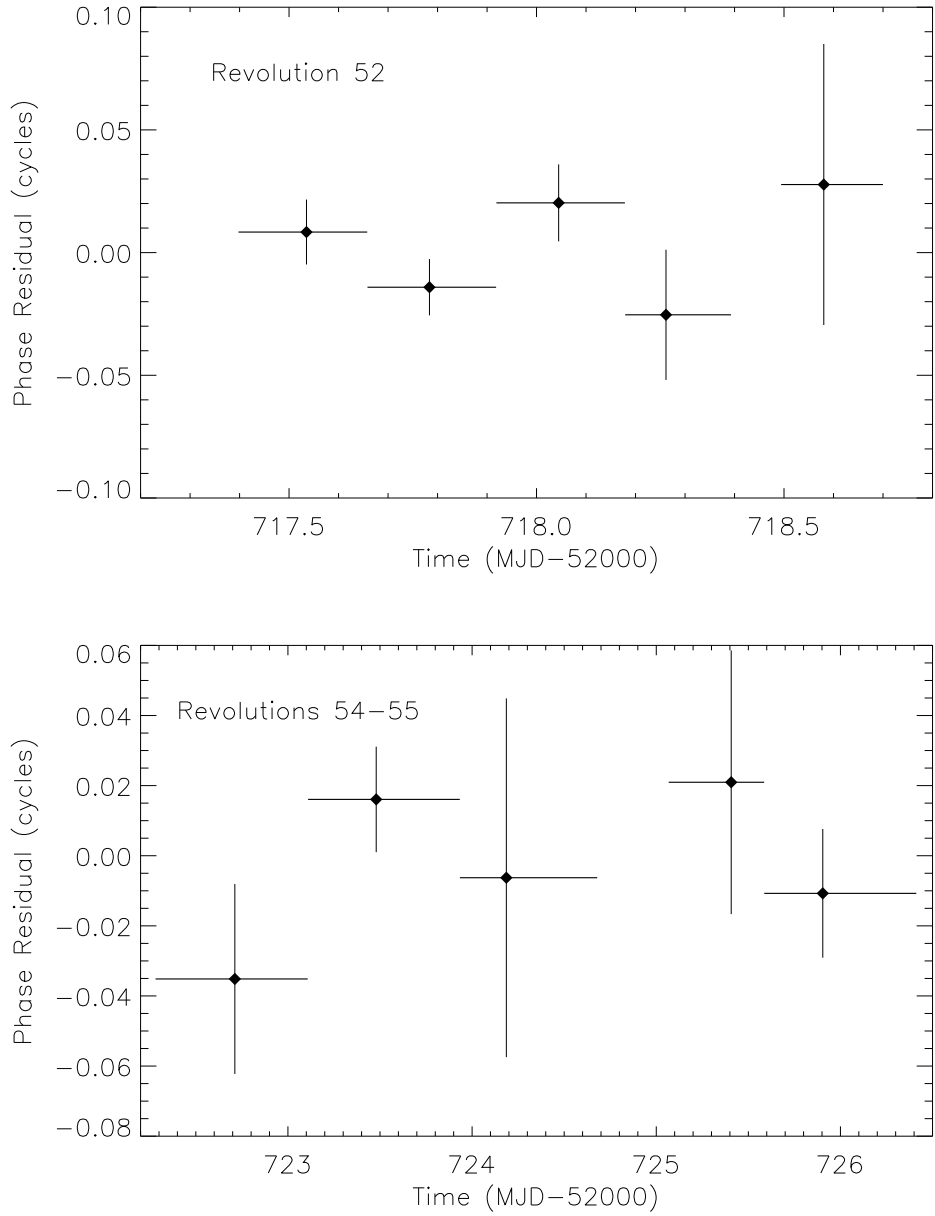}}
\end{center}
\caption{The phase offsets of the pulse profiles from sub-intervals of Rev 52 (top panel) and Revs 54$+$55 combined (bottom panel) relative to phase ephemerides with constant periods of 5964.7 s and 5870.9 s, respectively. \label{pulse_phases}}
\end{figure}

\begin{figure}[ht]
\begin{center}
\plotone{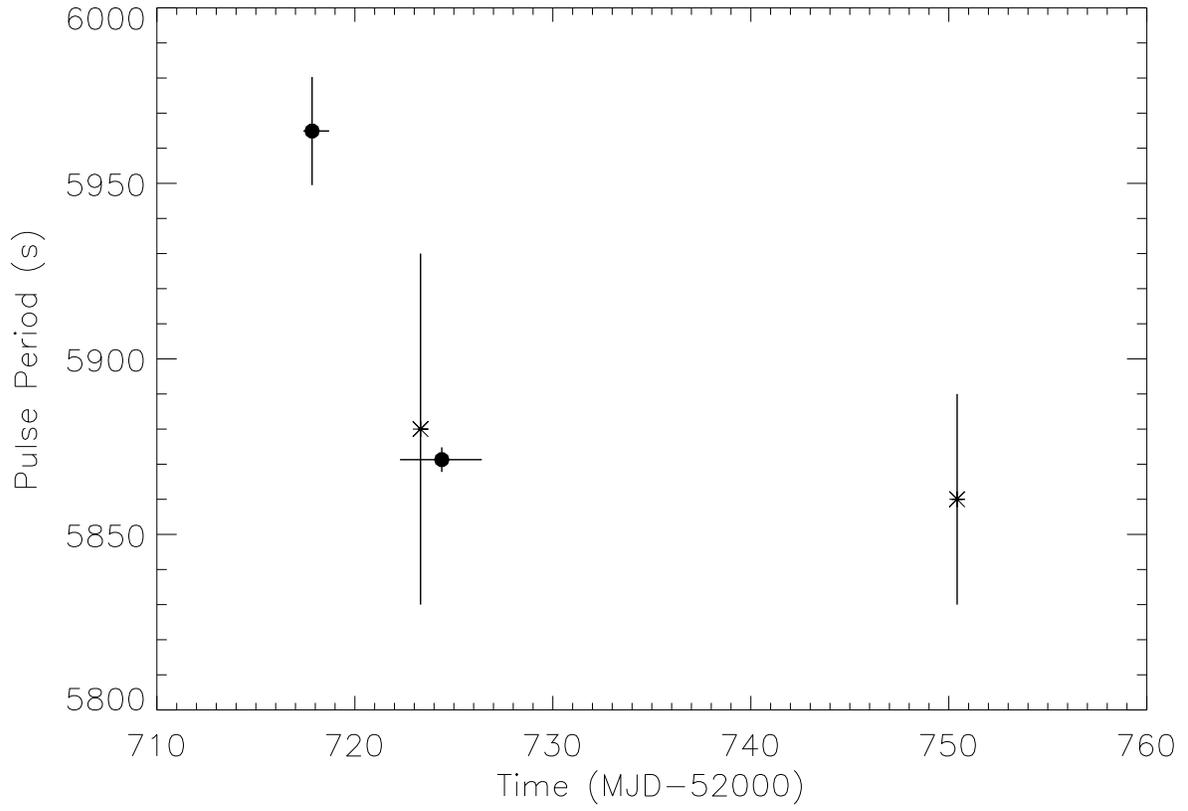}
\end{center}

\caption{The pulse periods determined from the ISGRI data ($\bullet$) along with the measurements based on Chandra observations ($\ast$).\label{pulse_periods}}
\end{figure}

\begin{figure}[ht]
\begin{center}
\scalebox{0.7}{\plotone{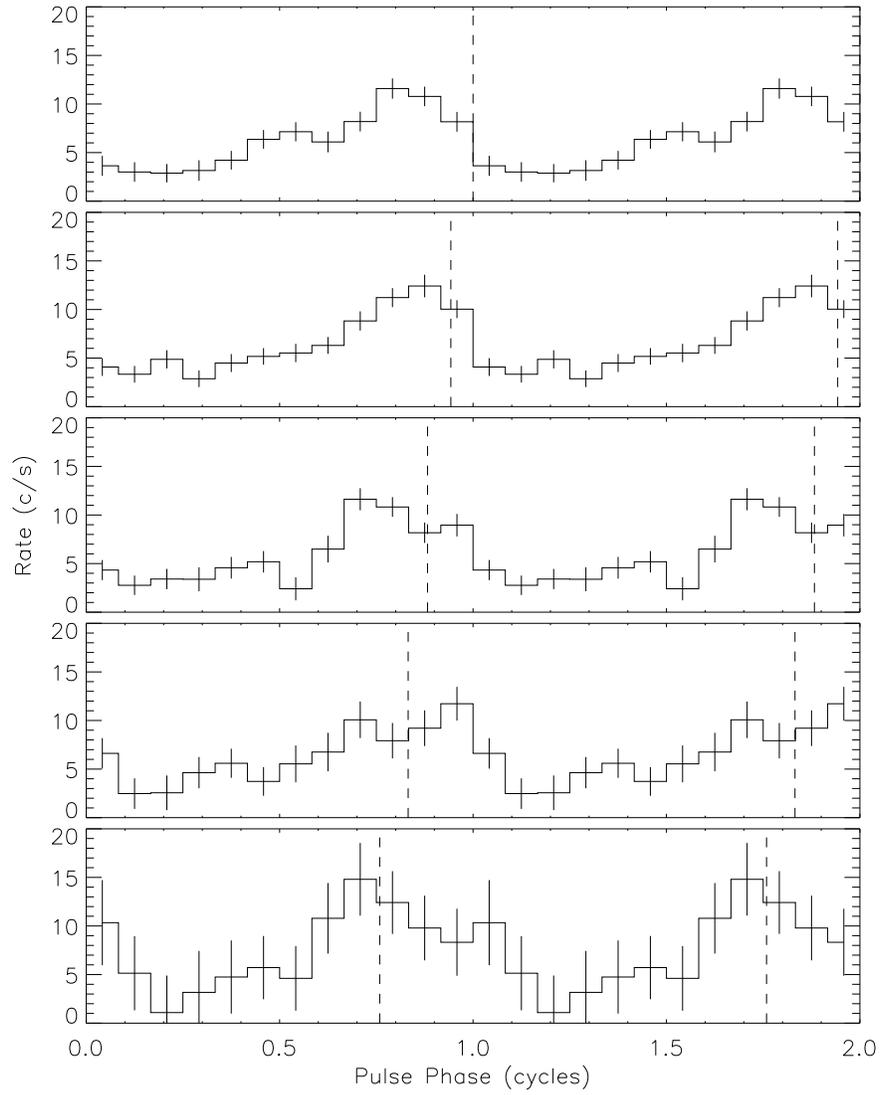}}
\end{center}
\caption{Pulse profiles for five sub-intervals in Revolution 52 obtained by epoch-folding the $20-45$ keV rates at a period of 5965~s. The dashed lines show the location of phase zero which would have resulted if the rates were folded with the 5871.3~s period determined for Revolutions 54$+$55.\label{profiles_52}}
\end{figure}

\begin{figure}[ht]   
\begin{center}
\plotone{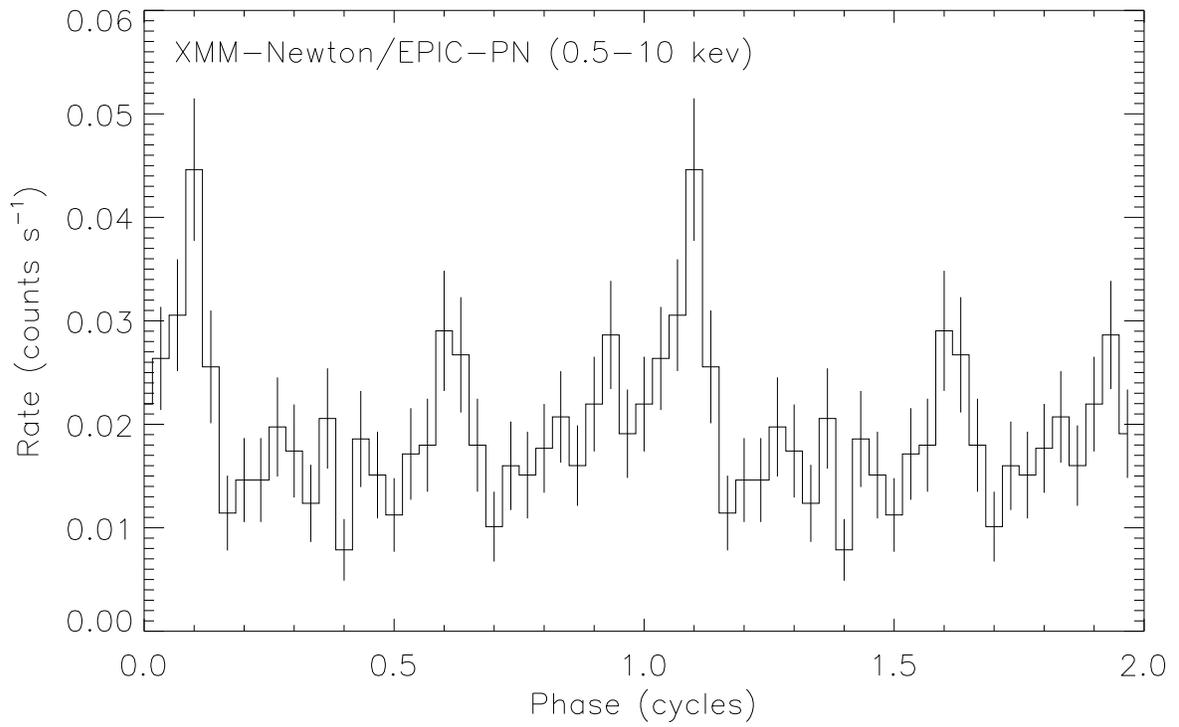}
\end{center}
\caption{Folded \xmm\ EPIC/PN pulse profile ($0.5-10$ keV) with period $P=5858\pm74$~s measured in 2004 September and referenced to the same epoch as in Figure \ref{prof}.}
\label{xmmprof} 
\end{figure} 

\begin{figure*}[ht]   
\begin{center}
\scalebox{0.8}{\plotone{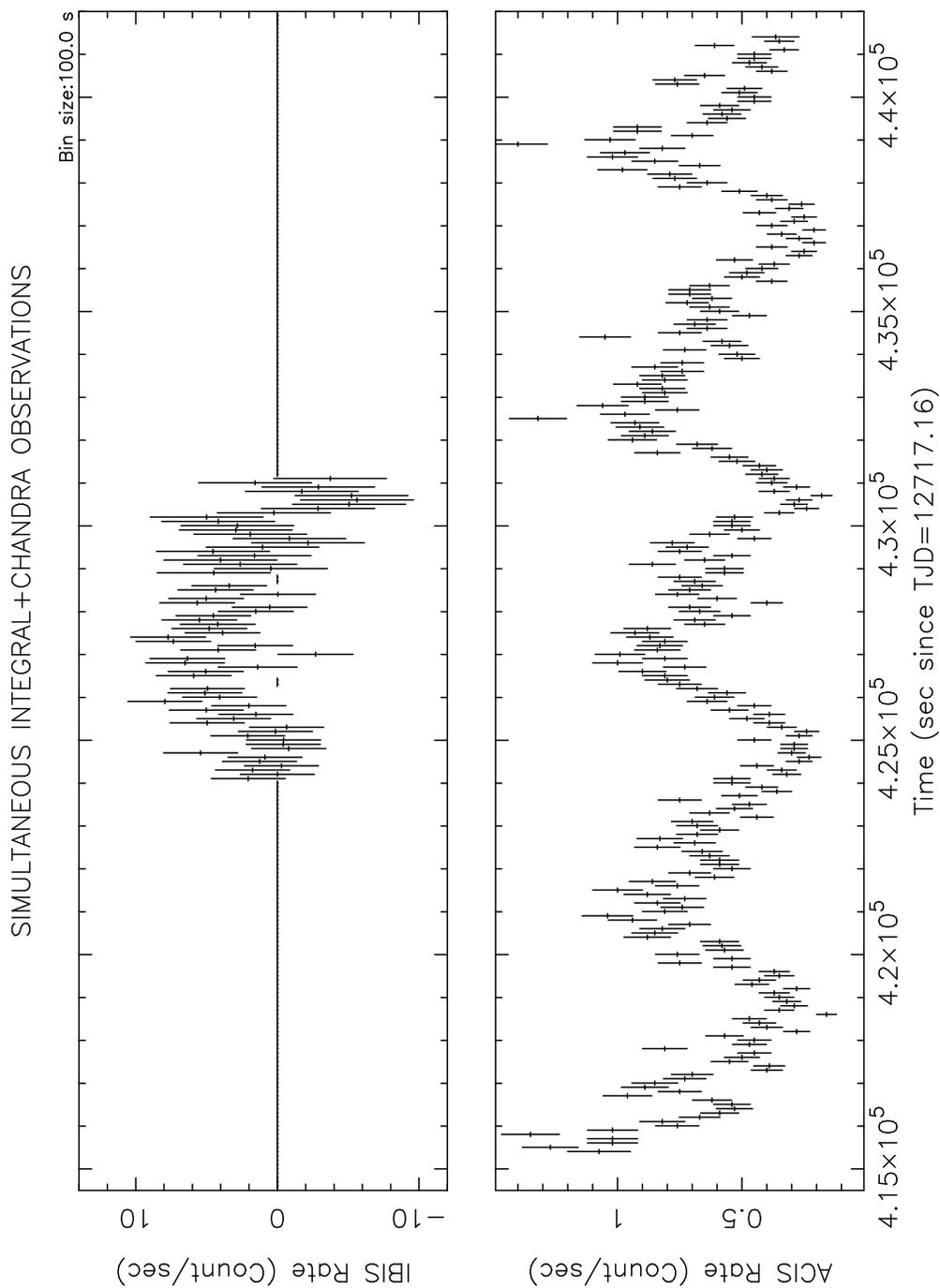}}
\end{center}
\caption{Simultaneous \chan\ ACIS-S2 ($2-10$ keV) and \integral\ ISGRI
  ($20-40$ keV) lightcurve of IGR~J16358$-$4726.} 
\label{jointlc} 
\end{figure*} 

\begin{figure}[ht]    
\begin{center}
\scalebox{0.65}{\plotone{f12.eps}}
\end{center}
\caption{Joint \chan\ ACIS-S2 and \integral\ ISGRI $\nu F_\nu$ spectrum of IGR~J16358$-$4726 during outburst on 2003 March 24. A highly absorbed Comptonization model gives the best fit to the data. A Fe K$_\alpha$ line feature clearly seen at 6.4~keV during the 24 March observation had faded below the detection capability of \chan\ during the second observation on 2003 April 21 \citep{Pat04}. The resisduals to the best fit reported in Table 4 are shown in the bottom panel.}
\label{jointspec} 
\end{figure} 

\begin{figure}[ht]   
\begin{center}
\plotone{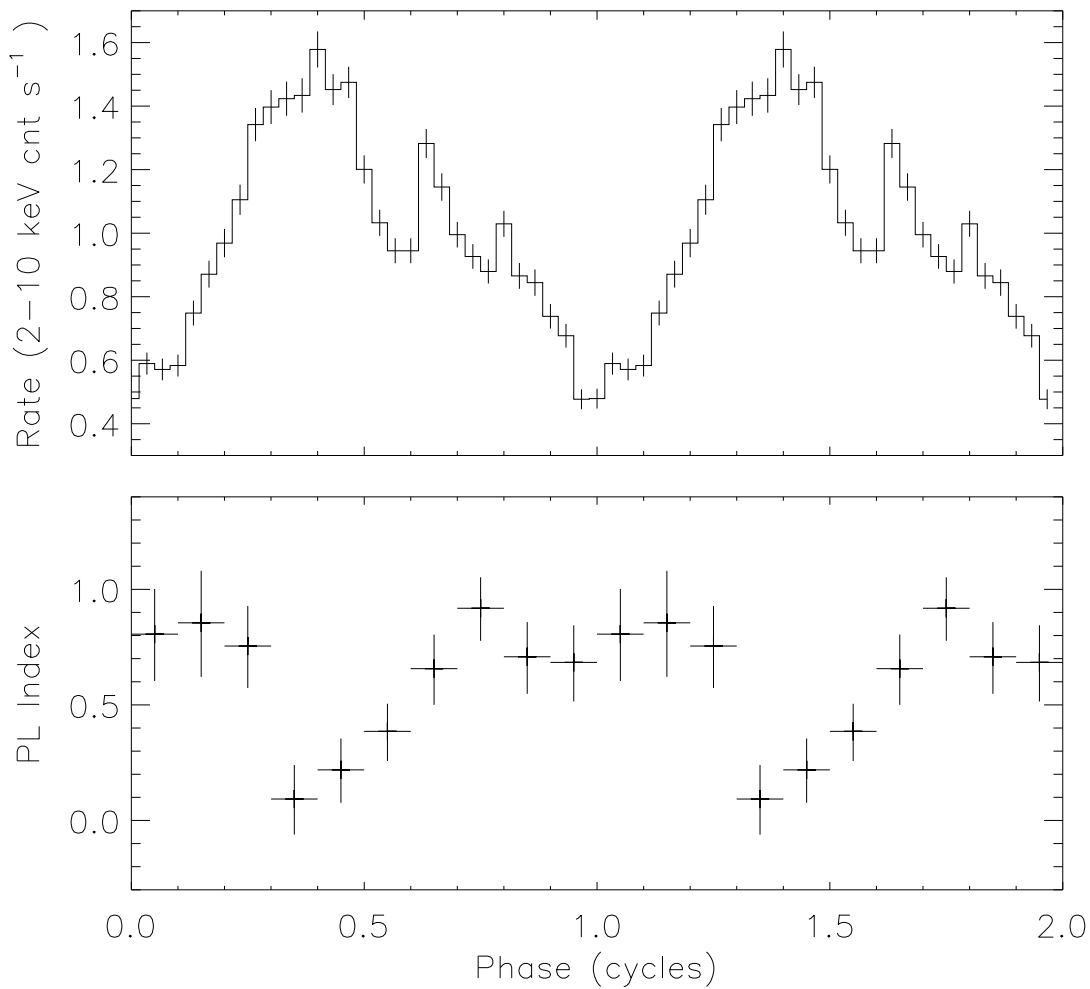}
\end{center}
\caption{Phase resolved spectral fit results derived with \chan\ ACIS-S2 (2.0$-$10.0~keV) data of IGR J16358$-$4726 in outburst on 2003~March~24. The top panel shows the 2-10 keV observed pulse profile (with 30 phase bins) and the bottom panel shows the spectral powerlaw index as a function of pulse phase (in 10 phase bins). There is evidence for evolution in the spectral index with the hardest spectra occurring near the end of the pulse rise and a clear spectral softening during the pulse decline. }
\label{phaseres}
\end{figure} 

\begin{figure}[ht]   
\begin{center}
\plotone{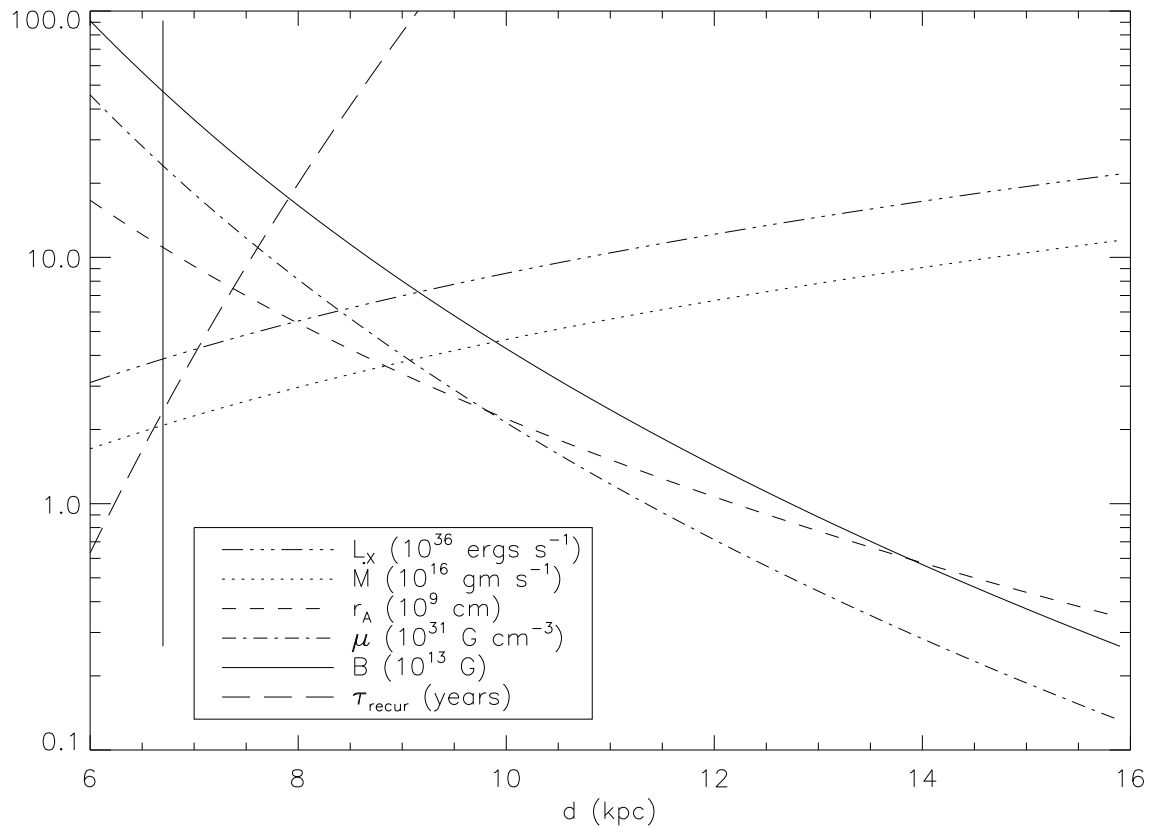}
\end{center}
\caption{The IGR~J16358-4726 2-100 keV luminosity, mass accretion rate, Alf\'{v}en radius, magnetic  moment, polar magnetic field strength, and recurrence time are plotted as a function of distance. The y-axis units for each curve are described in the legend.  The curves assume canonical neutron star mass, radius, and moment of inertia of $m_{1.4}=1$, $R_6 = 1$, and $I_{45} = 1$. The vertical solid line reflects the lower limit on the distance based on our 2004 XMM measurement.}
\label{accretionparameters}
\end{figure} 

\end{document}